\documentstyle[12pt]{article}

\begin{document}
\vspace{0.5cm}
\noindent
\baselineskip 20pt
\newcommand{\BE} {\begin{equation}}
\newcommand{\EE} {\end{equation}}
\newcommand{\BG} {\mbox{\rm\bf G}}
\newcommand{\BGO} {\mbox{${\rm\bf G}_0$}}
\newcommand{\DR} [1] {\raisebox{.2ex} {\mbox{$\stackrel {\rightarrow} {\rm\bf
D}_{#1}$}} }
\newcommand{\DL} [1] {\raisebox{.2ex} {\mbox{$\stackrel {\leftarrow} {\rm\bf
D}_{#1}$}} }
\newcommand{\lsim}{\raisebox{-3pt}{$\stackrel{<}{\sim}$}}
\newcommand{\gsim}{\raisebox{-3pt}{$\stackrel{>}{\sim}$}}

June, 1993
\hfill
TAUP   2059-93
\vspace{1.5 cm}

\centerline{\Large \bf
Bubble-Bubble Interaction in the}
\centerline{\Large \bf
Quark-Gluon Plasma}
\vspace{0.5 cm}
\centerline{by}
\vspace{0.5 cm}
\centerline{\bf Gideon Lana}

\bigskip\centerline{School of Physics and Astronomy}
\centerline{Raymond and Beverly Sackler Faculty of Exact Sciences}
\centerline{Tel Aviv University, Ramat-Aviv 69978, Israel}
\vspace{2.0 cm}

\centerline{\bf ABSTRACT}

We calculate the two-body interaction of spherical hadronic bubbles immersed in
a gluon plasma at temperatures above the phase transition.
Modeling the bubbles with the MIT bag we find that the two body potential is
repulsive for all bubble-bubble distances considered.
This implies that a static configuration of spherical hadronic bubbles in a
gluon background is consistent with the bag model of QCD.
\vfil\eject

\section{Introduction}

Considerable theoretical work \cite{PW,FJ,SY} and many computer simulations
\cite{BP,HP} of high temperature QCD have focused on the order of the phase
transition between quark-gluon plasma and hadronic matter. Quenched simulations
convincingly showed the transition is first order but weak i.e., with a
relatively small latent heat. When dynamical quarks are incorporated, the
transition weakens as smaller values of the quark mass are being used, and
recent calculations suggest that in the limit $m_q\rightarrow 0$ the transition
becomes second order \cite{SG}.

Most investigations modeling this transition in the context of the Universe
about 10-20 $\mu$sec after the big bang have assumed that the QCD transition
proceeds by mechanisms similar to those encountered in first order phase
transitions in condensed-matter systems, such as the condensation of water
vapor.
The common picture is of a homogeneous medium of QGP which cools adiabatically
as the universe expands; as the temperature drops just below $T_0$, which is of
the order of 100-200 MeV, the QGP becomes metastable, and hadronic bubbles
nucleate by local fluctuations. These seeds expand slowly while exchanging heat
and baryon number with the rest of the universe \cite{EW,KH,FA}.

Recently, however, an exact bag-model calculation suggested that the high
temperature phase of QCD is not a simple homogeneous medium of quarks and
gluons \cite{MS,BS}. It was claimed that at temperatures just above $T_0$ a
single bubble of finite radius of hadronic matter in the plasma has lower free
energy than the plasma with no bubbles at all. This means that the high
temperature phase of QCD may be composed of a plasma ``soup'' with floating
bubbles filled mostly by an ideal gas of pions. However, the evidence for this
phenomenon from lattice Monte-Carlo calculations  \cite{HPR,KPR,SO} is not
clear and the interpretation of the results is not too convincing.
Without addressing to the issue of how well the bag-model describes QCD physics
at high temperatures, we try to understand whether within the arena of the
bag-model such mixed-phase structure is indeed possible.

The bag model calculation \cite{MS} was performed for a single spherical
hadronic  bubble immersed in a homogeneous medium of plasma. The first question
which arose was whether this spherical configuration is indeed a minimum of the
free-energy functional and not a saddle point. In particular, is a spherical
bubble stable against small deformations of shape?
Using rather straightforward analysis it was shown that perturbations of its
boundary indeed tend to grow, suggesting that a spherical bubble may be
unstable \cite{LS}.

The aim of this work is twofold: to see if bubble-bubble interactions may, on
the one hand, support a finite density of bubbles, and, on the other hand,
stabilize the walls of the bubbles, such that the spherical bubbles discussed
in \cite{MS,LS} are indeed stable objects within the scope of MIT bag model.

The next section will be devoted to the explanation of the strategy of
obtaining an estimate of the bubble-bubble interaction.
In section \ref{secmre} we shall show how the multiple reflection expansion may
be applied to calculating the density of energy levels for the simplest
two-bubble geometry, and some details of the calculation will be given in
section \ref{secgeo}.
In section \ref{secres} the results of the calculation will be displayed and in
section \ref{secdis} we shall discuss the implications of the results and the
validity of the calculation.

\section{The Method}

We assume that the high temperature phase of QCD is composed of identical
spherical domains (bubbles) where the fields are manifested as colorless
excitations (mainly pions); the bubbles are surrounded by freely propagating
gauge fields.

In the early universe, the rate of expansion is much slower than the time
scales of QCD. This suggests that the field configuration is, at any time, in
complete thermal equilibrium and, specifically, that there are no temperature
gradients in the medium. Accordingly, we view it as a static configuration. We
further assume that the fluctuations in bubble-bubble distances are small.

Notice that in the absence of baryon number, the bubbles' radii and the  mean
distance between them completely specify the configuration. The structure we
have in mind is rather similar to that of a type II superconductor; the
hadronic domains form in our case a three dimensional lattice within the
quark-gluon plasma, whereas in the type II superconducting phase the geometry
is essentially two dimensional.

Since bubbles are color singlet objects, their interaction should stem from a
mechanism similar to the one producing the Casimir effect at $T=0$.
Such forces usually fall off very quickly as the relevant distance of the
problem is increased.
This leads us to assume that as long as the distance $d$ between the bubbles is
not too small, the interaction between bubbles is dominated by a two body
interaction, i.e., that three body and higher order contributions are
negligible in all configurations which are important to the partition function.
Upon completion of the calculation, a consistency check for this assumption
will be made.
The smallness of the interaction enables us to express the free energy of a
two-bubble configuration as
\BE
F_{bb}(R,d) = 2 F_b(R) + \delta\! F_{bb}(R,d) \, ,
\label{fbbdef}
\EE
where $F_b(R,d)$ is the free energy of a single bubble of radius $R$ in
infinite space, and $\delta\! F_{bb}(R,d)$ is a small correction to be derived.
For brevity of notation we omit the explicit $T$ dependence.
We first calculate the shift of energy level density $\delta\! \rho(k)$ for
each gluon, from which the free energy shift  $\delta\! F_{bb}(R,d)$ is derived
via
\BE
\delta\! F_{bb}(R,d)=n_g T\int_0^\infty\! d\!k \,\delta \! \rho(k)
\ln\left(1-e^{-k/T}\right)
\label{deltaf}
\EE
where $n_g=8$ is the number of gluons.

In the next section we shall briefly expound the multiple reflection expansion,
the tool with which $\delta \! \rho(k)$ is isolated and estimated.

\section{The Multiple Reflection Expansion}
\label{secmre}

The multiple  reflection expansion is a method of calculating the distribution
of eigenfrequencies for the wave equation in a connected finite domain of
arbitrary shape. It was developed for scalar and vector fields \cite{BBs,BBv}
and later for the much more involved case of Dirac fields \cite{HJ}.
It uses the time independent Green function formalism for obtaining an
expansion in $1/k$ where $k$ is the wavenumber.
The general idea is to extract the density of eigenfrequencies $\rho(k)$ from
the discontinuity of the propagator along the cut in the complex $k$ plane,
\BE
\rho_\gamma(k)= {2\over\pi} \left[ {\rm Im} \int_V \, d^3\! r\,  {\rm
Tr}\,\BG_\perp (r r^\prime,k) \right]_{r^\prime=r}\, ,
\label{rhok}
\EE
where $V$ is the volume in which the fields propagate. $\gamma$ is an imaginary
part added to $k$ which measures the width of a Lorentzian smearing function,
chosen in order to turn the discrete eigenvalue distributions that pertains to
finite domains into continuous ones. $\gamma$ must be chosen to be as small as
possible so that $k$ lies as close as possible to the real axis.

$\BG_\perp (r r^\prime,k)$ is the physical propagator which satisfies the
differential equation together with the boundary conditions. In our model the
fields satisfy the Helmholtz equation inside $V$ ,
\BE
(\nabla^2 + k^2)\vec A =  0 \, , \qquad {\rm div}\vec A=0 \, ,
\EE
and dual-superconductor boundary conditions on $S$, the boundary of the domain,
\BE
\hat n \cdot \vec E = \hat n \times \vec B = 0 \, .
\EE
As no confusion may arise, we shall henceforth omit the explicit $k$ dependence
of $\BG$.

For deriving expressions for the energy level density it is technically
convenient to augment the transverse propagator with a longitudinal part which
propagates longitudinal modes,
\BE
\BG=\BG_\perp+\BG_\parallel \, .
\EE
Using $\BG$ for calculating $\rho_\gamma(k)$ via Eq. (\ref{rhok}), this
additional piece yields an extra contribution which will be subtracted at the
end.
Proceeding along the lines of \cite{BBv}, we define for each point $\alpha$ on
the boundary of the domain the operators
\begin{eqnarray}
\DL{\alpha}& = & {\bf P}_n(\alpha) + {\stackrel{\leftarrow}{\partial}
  \over \partial n_\alpha} {\bf P}_t(\alpha) \, , \nonumber \\
\DR{\alpha} & = & {\bf P}_n(\alpha) {\stackrel{\rightarrow}{\partial}
  \over \partial n_\alpha} - {\bf P}_t(\alpha) \, ,
\end{eqnarray}
where ${\bf P}_n(\alpha)$ and ${\bf P}_t(\alpha)$ are projection operators on
the normal $\vec n_\alpha$ and on the tangent plane to the surface at the point
$\alpha$ respectively.
Augmenting also the free transverse propagator with a longitudinal component,
\BE
\BGO=\BGO_\perp+\BGO_\parallel={\mbox{\bf 1}} \, G_0 \,  ,
\qquad
G_0={\exp(ik|r-r^\prime|) \over 4\pi\, |r-r^\prime|} \,  ,
\EE
enables us to expand $\BG$ in terms of $\BGO$, yielding a multiple reflection
expansion:

\begin{eqnarray}
\BG(r r^\prime)& = &\BGO(r r^\prime) + 2 \int_S d\sigma_{\!\alpha} \, \BGO (r
\alpha) \DL{\alpha}\DR{\alpha} \BGO (\alpha\, r^\prime) \nonumber \\
& &  + 2^2 \int_S d\sigma_{\!\alpha}\, d\sigma_{\!\beta} \,
\BGO
(r \alpha) \DL{\alpha}\DR{\alpha} \BGO (\alpha\beta)
 \DL{\beta} \DR{\beta} \BGO (\beta r^\prime)
+ \cdots \, .
\label{gmre}
\end{eqnarray}

The first term gives for the density of states the usual contribution
proportional to the (infinite) volume $V$, and is therefore irrelevant to
$\delta \! \rho(k)$ .
When computing the free energy of the system it is subtracted off.
The second term, which corresponds to a single reflection, vanishes identically
upon performing the trace and subtracting the contribution of the longitudinal
modes.

For the two-reflection term the integrand is
\begin{eqnarray}
& & G_0(r \alpha) {\partial G_0(\alpha \beta)\over\partial n_\alpha}
{\partial G_0(\beta r^\prime) \over\partial n_\beta} (1-\sin^2 \theta)
\nonumber \\
&& - \Biggl[G_0(r \alpha) {\partial^2 G_0(\alpha \beta)\over\partial n_\alpha
\, \partial n_\beta} G_0(\beta r^\prime)
+ {\partial G_0(r \alpha) \over\partial n_\alpha}  G_0(\alpha \beta) {\partial
G_0(\beta r^\prime) \over\partial n_\beta}
\Biggr]\sin^2 \theta \nonumber \\
&&+ {\partial G_0(r \alpha) \over\partial n_\alpha}
 {\partial G_0(\alpha \beta)\over\partial n_\beta} G_0(\beta r^\prime)
(2-\sin^2 \theta) \, ,
\label{g_two_refl}
\end{eqnarray}
where $\theta$ denotes the angle between the normal vectors to the surface $S$
at the points $\alpha$ and $\beta$, namely $\cos\theta=n_\alpha\cdot n_\beta$.
Upon integrating over $\alpha$, $\beta$ and taking $r=r^\prime$ the triple
products of propagators in the first and last terms give identical results.
Collecting all terms, subtracting the contribution of the longitudinal modes,
and performing the trace, we finally get for the two-reflection shift in the
density of states
\begin{eqnarray}
 \delta\! \rho_\gamma(k) & = & {8k\over \pi} {\rm Im} \int_V d^3\!r \int_S
d\sigma_{\!\alpha}\, d\sigma_{\!\beta} \,
\Biggl[2 G_0(r \alpha) {\partial G_0(\alpha \beta)\over\partial n_\alpha}
{\partial G_0(\beta r) \over\partial n_\beta}  \nonumber \\
& & -\sin^2 \theta {\partial^2\over\partial n_\alpha \, \partial n_\beta}
\left( G_0(r \alpha)  G_0(\alpha \beta) G_0(\beta r) \right)\Biggr] \, .
\label{trg}
\end{eqnarray}

\section{The Geometrical Setting}
\label{secgeo}

In order to calculate the two-bubble interaction alone, it is sufficient to
consider a simple geometry of two spherical bubbles in infinite space.
This is in practice similar to confining the two bubbles in a cavity, and
taking the walls of the cavity to infinity.
The actual application of this limiting procedure is considerably simplified by
the fact that only the bubble-bubble interaction is of interest and not the
artificial interactions of the bubbles with the walls of the cavity. Therefore,
we do not have to take into account reflections from the walls, i.e., the
integrations over the points $\alpha$ and $\beta$ are performed only over the
two spherical surfaces of the bubbles.

Figure \ref{fig_twoterms}a shows schematically the relevant contribution to the
propagator $\BG$, where the limit $r\rightarrow r^\prime$ has already been
taken. The points $\alpha$ and $\beta$ are on different spheres and $r$ is in
$V$.

For the free energy shift $\delta\! F_{bb}$ of the two-bubble configuration
relative to the free energy of two completely separate bubbles, one has to
subtract an excluded-volume contribution.
Had we computed the free energy of a single bubble in infinite space by using
the multiple reflection expansion, the two-reflection term would be as depicted
in Fig. \ref{fig_twoterms}b: The points $\alpha$ and $\beta$ lie on the surface
of the bubble, and the coordinate $r$ runs over all space, the interior of the
single bubble excluded.
When introducing another bubble, its interior is no longer included in the
region of integration, and therefore the excluded-volume terms must be
subtracted off.

\begin{figure}
\vspace{6.0 cm}
\caption{Schematic representation of the two kinds of contributions to the two
reflection term. See text.}
\label{fig_twoterms}
\end{figure}

\section{Results}
\label{secres}

We have computed the shift in the density of states $\delta\!\rho_\gamma(k)$ as
a function of the common radius $R$
of the bubbles and the distance $d$ between their centers.
This enables us to estimate the free energy shift associated with the
interaction of bubbles by Eq. (\ref{deltaf}).

Since for the equilibrium bubble radius (see Refs. \cite{MS,BS}) $RT\simeq 1$ ,
it was sufficient to compute $\delta\!\rho_\gamma(k)$ in the range $0.2\leq
kR\leq 10$. Lower or higher values of $k$ are expected to yield too small
contributions in Eq. (\ref{deltaf}).
The convergence parameter $\gamma$ was chosen as small as the numerics allowed.
The results which follow correspond to the choice of $\gamma=0.4/R$.
Calculations for lower values of $\gamma$ require much more iterations to
achieve similar accuracy, but lead substantially to the same results.
Physically, such a high value of $\gamma$ may be viewed as effectively
introducing a gluon mass into the calculation.

Integrating Eq. (\ref{trg}) we encounter seven-dimensional integrals.
Exploiting the azimuthal symmetry reduces them to six-dimensional integrals. We
were, therefore, forced to resort to numerical routines which sample the huge
space. We used the computer code VEGAS \cite{V}, which we found to be very
efficient for the task..
We ran the numerical program on a Cyber 920 Silicon-Graphics machine. For each
value of $d$ the program ran for about 30 hours, sampling about $2\cdot 10^6$
points for each value of $k$.

The integration was performed using the bispherical coordinate system (see e.g.
\cite{MF}) in terms of which $S$ is trivially expressed as a union of the
surfaces $\mu=-\mu_0$ and  $\mu=\mu_0$ where  $\mu$ is the radial coordinate;
$\mu_0$ fixes the ratio between the bubbles' radii and separation via
$d/2R=\cosh \mu_0$. The volume integration is then performed over the range
$-\mu_0< \mu< \mu_0$ for (a) type terms, and $\mu_0< \mu$ for (b) type terms
(see Fig. \ref{fig_twoterms}).

In general, the limit $r\rightarrow r^\prime$ must be taken with caution, since
for a simply connected boundary all four points $\alpha$, $\beta$, $r$ and
$r^\prime$ may become arbitrarily close.
In our case, however, the boundary is not simply connected and for both (a) and
(b) type terms, the four points may not coincide.
Therefore, the superficial singularities of the integrand of (\ref{trg}),
$\alpha\rightarrow r$ or $\beta\rightarrow r$ for (a) type terms, and
$\alpha\rightarrow \beta$ for (b) type terms, are not actual singularities due
to the vanishing of the measure.
 From the numerical point of view, it is then sufficient to avoid sampling the
space of integration too close to subspaces where the value of the integrand of
(\ref{trg}) becomes unbounded.

Figure \ref{figrho} is a typical plot of the shift in the density of levels as
a function of $k$ for $R=1$ fm and $d=2.59$ fm. As expected, for small values
of $k$,  $\delta\! \rho(k)$ approaches zero. This feature is common to all
combinations of $R$ and $d$ we explored.
The numerical integration, however, fails to yield the correct decrease of
$\delta\! \rho(k)$ for large values of $k$, where the integrand is highly
oscillatory. The insufficiency of the sampling mesh for large values of $k$ is
also reflected in the large error bars, which are of the order of the estimate
itself. This does not prevent us from obtaining a reliable estimate for
$\delta\! F(k)$ since for large values of $k$ the logarithmic factor in Eq.
\ref{deltaf} is exponentially small.

\begin{figure}
\vspace{8.0 cm}
\caption{Typical behavior of $\delta\! \rho(k)$ for fixed values of $R$ and
$d$. }
\label{figrho}
\end{figure}

The free energy shift as a function of the bubbles' radius and their distance
is shown in Fig. \ref{figfdr}. The relative errors, which are not shown here,
are typically around 10 to 30 percent, though for the largest values of $d/R$
where $\delta\! F$ goes to zero the relative errors are higher.
The most important feature is the increase of $\delta\! F$ for small values of
$d/R$ which is most pronounced for $0.5\, {\rm fm}\lsim R\lsim 2\, {\rm fm}$.
The meaning of this is that bubbles tend to repel each other, i.e., bubbles do
not tend to coalesce. This result supports the picture suggested in Refs.
\cite{MS,BS} of a phase of quark-gluon plasma populated with hadronic bubbles
arranged in a lattice-like structure.

Another important result is that the largest values obtained for $\delta\! F$
are of the order of $100$ MeV, and typical values are of the order of $10$ MeV.
This is to be compared with the free energy shift $F_b$ due to a single bubble
[see  Eq. (\ref{fbbdef})], which is typically about $500$ MeV at the
equilibrium radius. This means that the presence of neighboring bubbles should
have little effect on the equilibrium radius.
The mean separation between bubbles, however, is determined both by $F_b$ and
$\delta\! F_{bb}$: On the one hand, the tendency of $F_b$ is to populate the
plasma with bubbles as densely as possible. On the other hand, $\delta \! F$
pushes them away from each other as far as possible. The net result of these
two mechanisms may be a stable lattice of bubbles.

\begin{figure}
\vspace{16.0 cm}
\caption{$\delta\! F$ for as a function of $R$ and $d/R$ for a temperature
$T=155 MeV$.}
\label{figfdr}
\end{figure}

Three limiting regimes are worth mentioning.
 For $d-2R\ll R$ the results of the calculation are not shown, since for this
case the distance between the surfaces of the bubbles is very small, and the
propagator $\BGO$ will decrease considerably only after many reflections. In
other words, we expect that many reflections should be important in such
situation and the two-reflection term badly approximates  $\delta\! F$.
 For the limit $R\rightarrow 0$ with $d$ fixed, the shift in the level density
tends to zero due to the vanishing of the surface $S$ in the integral
(\ref{gmre}).
 For $d/R\rightarrow \infty$ the vanishing of $\delta\!$ is clearly exhibited;
this is to be expected due to the $1/R$ behavior of the propagator.

We are particularly interested in the behavior of $\delta \! F(R,d)$ as a
function of $d$ for fixed $R$.
For each temperature we fix the value of $R$ to the radius that minimizes the
free energy for a single bubble geometry (see Ref. \cite{MS}). The cuts through
the free-energy surfaces yield the curves shown in Fig. \ref{figfd}.

\begin{figure}
\vspace{10.0 cm}
\caption{$\delta\! F$ for plotted as a function of $d$, the distance between
the bubbles for various temperatures. For each temperature $R$ is fixed to the
value which minimizes $F_b(R,t)$.}
\label{figfd}
\end{figure}

\section{Discussion}
\label{secdis}

We have estimated the shift of free energy related to the interaction between
two spherical bubbles of hadronic phase immersed in a gluon plasma. In the
relevant regime of parameters, a repulsion between the bubbles was found.
If quarks play only a minor role in this phenomenon, as was found in the single
bubble calculation in Ref. \cite{MS}, this finding supports the picture of the
Swiss cheese instability \cite{BS} of the early Universe.
As an approximation we have used the first non-trivial contribution of the
multiple reflection expansion. Since this is essentially an asymptotic
expansion in $1/k$, it is not easy to predict a priori the validity of such
truncation.

At first glance, truncating all higher-order reflections has some similarity to
the dilute gas approximation: For large $d$ the free propagator $\BGO$
decreases strongly for every trip from one bubble to another. In other words,
when the bubbles are dilute only terms which correspond to a minimal number of
trips between bubbles are important.
This may be misleading, since such terms arise at any higher order of the
expansion.

The correction we have computed is indeed considerably smaller than $F_b$, the
free energy of a single bubble in infinite space. But an explicit comparison
with terms stemming from larger number of reflections is completely
impractical, because of technical problems in calculating integrals of more
than nine dimensions. One may, however, be able to compute successive terms by
using completely different ways of performing this calculation: For example,
one might iterate the integral equation II.29 of Ref. \cite{BBv} on a grid, and
use this as an input for equation II.23.

Our results are only qualitative in nature; The error bars are considerable,
and we have only a limited window in $k$ from which we may get reliable
numbers. Therefore it seems to us premature even to use them to predict the
structure of the lattice of bubbles we suggest.

The repulsion we found, is intuitively exerted by a positive outside pressure
on the walls of the bubbles. However, by regarding the free energy as a
function of the bubbles' radii for fixed $d$, it may be immediately noticed
that it contains not only a pure pressure term, cubic in $R$, but other terms
as well. These terms are crucial to the issue of the surface instability of the
bubble, namely its tendency to grow fingerlike structures, as suggested in
\cite{LS}.
Again, much more precise calculation has to be performed in order to answer
whether the two-bubble interaction may eliminate or reduce the surface
instability.

\section*{Acknowledgments}
We thank Benjamin Svetitsky for many helpful discussions and for thorough
reading of the manuscript.
This work was supported by a Wolfson Research Award administered by the
Israel Academy of Sciences and Humanities.

\end{document}